\documentclass[%
 reprint,
 amsmath,amssymb,
 aps,
]{revtex4-2}

\usepackage{graphicx}
\usepackage{dcolumn}
\usepackage{bm}

\usepackage[colorlinks,citecolor=blue]{hyperref}

\usepackage[dvipsnames]{xcolor}
\usepackage{url}

\begin{document}

\preprint{APS/123-QED}

\title{Neural networks for on-the-fly single-shot state classification}

\author{Rohit Navarathna$^{1,2}$}
\email{r.navarathna@uq.edu.au}
\author{Tyler Jones$^{1,2,3}$}
\author{Tina Moghaddam$^{1,2}$}
\author{Anatoly Kulikov$^{1,2,4}$}
\author{Rohit Beriwal$^{1,2}$}
\author{Markus Jerger$^{1,2,5}$}
\author{Prasanna Pakkiam$^{1,2}$}
\author{Arkady Fedorov$^{1,2}$}
\affiliation{
 $^1$ ARC Centre of Excellence for Engineered Quantum Systems, St Lucia, Queensland 4072, Australia\\
 $^2$ School of Mathematics and Physics, University of Queensland, St Lucia, Queensland 4072, Australia\\
 $^3$ Max Kelsen, Spring Hill, Queensland 4000, Australia\\
 $^4$ Department of Physics, ETH Z\"{u}rich, CH-8093 Z\"{u}rich, Switzerland\\
 $^5$ JARA-FIT Institute for Quantum Information, Forschungszentrum J\"{u}lich, 52425 J\"{u}lich, Germany
}
\date{\today}

\begin{abstract}

Neural networks have proven to be efficient for a number of practical applications ranging from image recognition to identifying phase transitions in quantum physics models. In this paper we investigate the application of neural networks to state classification in a single-shot quantum measurement. We use dispersive readout of a superconducting transmon circuit to demonstrate an increase in assignment fidelity for both two and three state classification. More importantly, our method is ready for on-the-fly data processing without overhead or need for large data transfer to a hard drive. In addition we demonstrate the capacity of neural networks to be trained against experimental imperfections, such as phase drift of a local oscillator in a heterodyne detection scheme.

\end{abstract}

\maketitle


Machine learning (ML) is ubiquitous in modern computer science with a wide range of applications. By virtue of its archetypal problem classes -- regression and classification -- ML algorithms and neural networks in particular have recently found a number of applications in quantum computing, helping researchers to tackle such tasks as optimizing gates and pulse sequences \cite{Zahedinejad2016,August2017,Ding2021,Baum2021}, identifying phase transitions \cite{Rem2019, Dong2019}, correcting imperfections of measurement apparatus \cite{Palmieri2020, Zwolak2020,Durrer2020}, classifying states~\cite{Hueaav2761,Cimini2020} or evolution~\cite{Stenberg2016,Flurin2020,Gentile2021} of a quantum system with little or no {\it a priori} knowledge, and even optimizing the fabrication process \cite{Mei2021a}.

A proposal to use machine learning to discriminate measurement trajectories was one of the first and most natural applications of ML in the field, and has led to improvements in readout assignment fidelity \cite{Magesan2015}. The technique is now being regularly implemented across the community \cite{Dickel2018,Kono2018,Martinez2020} due to their generalisability and capacity to extract useful features from dense data. A recent advancement uses neural networks to compensate for system dependent errors due to processes such as cross-talk in multiplexed qubit readout~\cite{Lienhard2021}.  
In this work we also apply neural networks to the readout of a superconducting transmon system. However, our approach works on-the-fly with no data processing overhead and can be trained against experimental parameter drifts.

To deploy our neural-network-based state classification, we use an open source PyTorch library~\cite{NEURIPS2019_9015}. Geared towards computer vision and natural language processing, it includes the capability to realise deep neural networks and contains built-in functionality for data processing on a graphics processing unit (GPU). GPU integration enables our pipeline to be fast enough to perform on-the-fly data classification without the need to transfer raw measured signal to a hard drive. Amongst other advantages, it allows monitoring the readout assignment fidelity in real time. 

With the initial training of the neural network taking on the order of minutes, consequent retraining of the network weights requires several seconds and allows the readout assignment fidelity to return to the optimal value. More importantly, the convolutional neural network used in the present work may be designed and trained in a way resilient to certain experimental parameter drifts. Specifically, we present a strategy to eliminate the effect of local relative phase drifts induced by generating microwave equipment on the readout assignment fidelity.

\begin{figure}[hbtp]
\centering
\includegraphics[width=\linewidth]{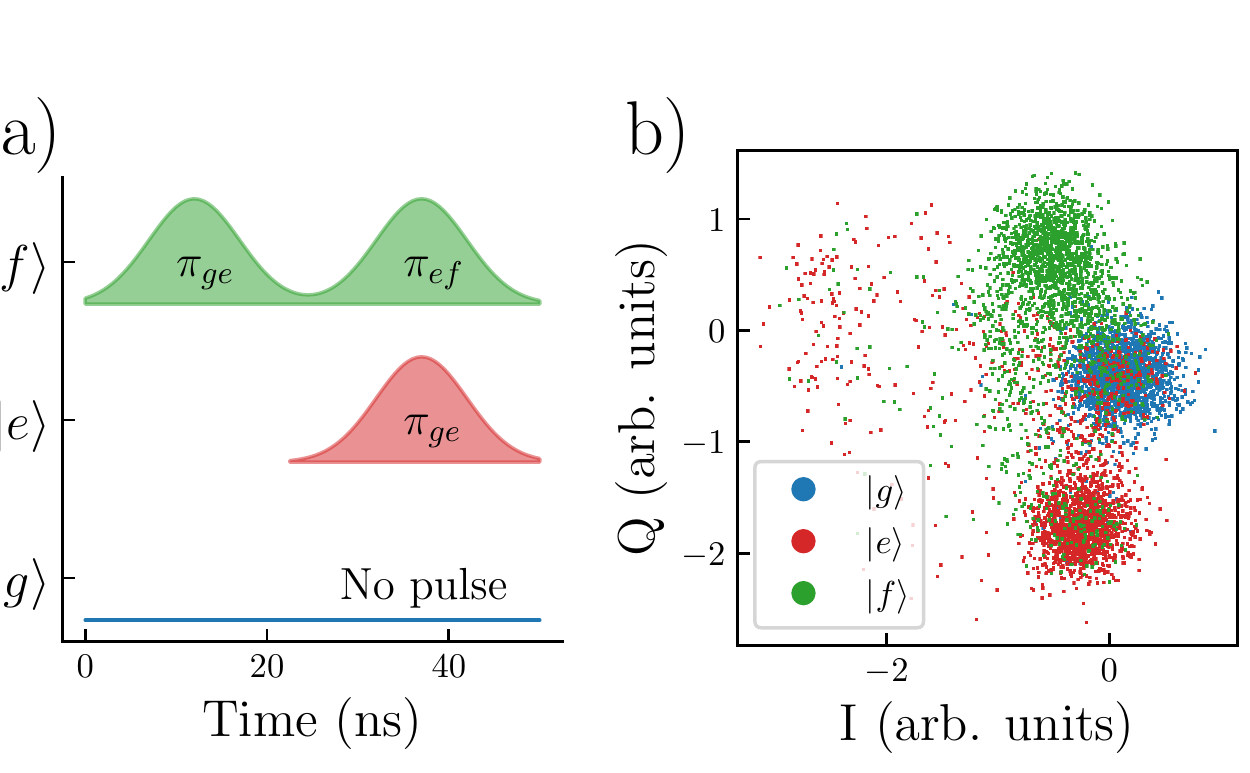}
\caption{a) Transmon control pulses used for preparing three basis states of the transmon. b) Histogram of integrated cavity responses $I+iQ$. The blue, red and green points correspond to the transmon being prepared in $\left|g\right>$, $\left|e\right>$ and $\left|f\right>$ states, respectively. The relaxation of the f-state to the excited state and excited state to ground state is visible. Some of the points that are decaying from the f-state to the excited state will be classified as ground state due to their proximity to the ground state cluster.}
\label{fig:blobs}
\end{figure}

In our experiment we used a primitive of the circuit quantum electrodynamics platform: a transmon coupled to a readout cavity. After initializing the qubit in its ground state, we used Gaussian pulses of length $20$~ns to prepare the transmon in the three basis states; applying no pulse to keep the transmon in its ground state $\left|g\right>$, applying one $\pi$-pulse at $\omega_{ge}$ to prepare the transmon in its first excited state $\left|e\right>$, and applying two consecutive $\pi$-pulses at $\omega_{ge}$ and $\omega_{ef}$ respectively to prepare a transmon in the second excited state $\left|f\right>$. These protocols are illustrated in Fig.~\ref{fig:blobs}a. 

After the state preparation we perform measurement by acquiring the probe signal transmitted through the readout resonator. To achieve the single-shot regime, we used a Josephson parametric amplifier (JPA)~\cite{Eichler2014}. Following amplification through the JPA, the readout signal is further amplified using a High Electron Mobility Transistor (HEMT) amplifier and multiple room temperature amplifiers. The signal is then downconverted to 25~MHz and acquired by a digitizer.
After repeating the protocol, acquiring the signal and classifying the states we calculate the assignment fidelity which was used to evaluate the efficacy of different classification methods. We performed the experiments with two different samples  (see Table~\ref{tab:device_parameters}). The first run (Sample A) of experiments was used as a test-bed to compare the quality of various ML methods. The second set of experiments (Sample B) demonstrated on-the-fly data processing using a GPU and the methods' stability against phase drifts. 

Within our data processing workflow, each acquired waveform undergoes digital downconversion (DDC) by multiplying the acquired signal with $\cos(\omega_{\rm DDC}t$) ($\sin[\omega_{\rm DDC}t]$) where $\omega_{\rm DDC}/2\pi = 25~$MHz, to obtain the in-phase  quadrature $I(t)$  (out-of-phase quadrature $Q(t)$). A Finite Impulse Response (FIR) Filter with a window of $40$ samples ($20$ ns) and a cutoff frequency of $20$ MHz is applied to the signal to eliminate the signal image at $50$~MHz along with $25$~MHz noise (originally DC offset). After obtaining $I(t)$ and $Q(t)$, the signal undergoes further post-processing. This may include time integration, channel correlation, or even being fed through trained PyTorch neural networks.

For the on-the-fly experiments with Sample B we acquired 512 time points per measurement, recorded to the buffer of a 500 MSa/s digitizer Spectrum M4i. After populating the buffer with 2048 time traces, we transferred data to PC memory (RAM) and then to the GPU memory for batch processing. While the data is being processed, the digitizer buffer is populated with new acquired waveforms. Due to the large number of cores in the GPU the data can be processed in parallel, which allowed us to perform real time data acquisition on-the-fly. Although the results of this paper were obtained with a repetition time of $40~\mu$s, our GPU data processing can be run without overhead as fast as $3.2~\mu$s per repetition, obtaining $19$ million traces ($19$ trillion samples) in $1$ minute.

\begin{table}
\begin{ruledtabular}
\begin{tabular}{lcdr}
\textrm{Parameter}&
\textrm{Transmon A} &\textrm{Transmon B}\\
\colrule
$\omega_{cav}/2\pi$  &7.08~\rm{GHz} &7.63~\rm{GHz}\\
$\omega_{ge}/2\pi$   &6.27~\rm{GHz} &5.49~\rm{GHz}\\
$\omega_{ef}/2\pi$   &5.95~\rm{GHz} &5.16~\rm{GHz}\\ 
$2\chi_{ge}/2\pi$    &8.00~\rm{MHz} &8.50~\rm{MHz}\\
$2\chi_{ef}/2\pi$    &5.35~\rm{MHz} &15.57~\rm{MHz}\\
$\kappa/2\pi$        &1.31~\rm{MHz} &1.56~\textrm{MHz}\\
$T_1$                &11.75~\rm{$\mu$s} &4.07~\textrm{$\mu$s}\\
$T_2$                &3.17~\rm{$\mu$s} &4.29~\textrm{$\mu$s}
\end{tabular}
\end{ruledtabular}
\caption{Device parameters. Here, $\omega_{cav}$ is the frequency of the readout resonator, $\omega_{ge}$ ($\omega_{ef}$) is the frequency of transition between the ground (first excited) state and the first (second) excited state, $\chi_{ge}$, $\chi_{ef}$ are the state-dependent dispersive shifts of the resonator frequency, $\kappa$ is the decay rate of the resonator and $T_1$ ($T_2$) is the relaxation (dephasing) rates of the transmons.}
\label{tab:device_parameters}
\end{table}

In order to determine the {\it baseline} readout fidelity, we first employed the conventional classification. Following this method, the heterodyne measurement signal was integrated in time, giving us one complex number $I$ and $Q$. By repeating the measurements, we can populate histograms for every prepared state on the I-Q plane, as shown in Fig.~\ref{fig:blobs}. A measurement response can be classified by selecting the state whose mean response is closest to the mean state response on the I-Q plane. The assignment fidelity can be evaluated as $\mathcal{F}_a = (1/N) \sum_{i=1}^N \mathbb{P}(i|i)$, where $\mathbb{P}(i|j)$ is the probability of obtaining outcome “$i$” given the system was prepared in $j^{\textrm{th}}$ state. Here, we use $N = 2 (3)$ and states ${\left|g\right>, \left|e\right>}$ (${\left|g\right>, \left|e\right>,\left|f\right> }$) to calculate qubit (qutrit) assignment fidelity.

Alternatively, one can apply a {\it matched filter} to the heterodyne measurement signal prior to integration in the conventional method. Matched filters are calibrated by taking the means of all acquired signals corresponding to each basis state. An incoming signal is convolved with  these filters. The filter which returns maximum average amplitude determines the classified basis state. 

To identify the best ML algorithm  we first collected data from Sample A. The transmon was prepared in the three basis states followed by the 2~$\mu$s measurement pulse, resulting in 50 time samples for each trace at 100~\rm{MHz}. In total, we collected 16384 traces corresponding to each state for analysis. $90\%$ of this data was used for training, and the rest was used to test and obtain the assignment fidelity.

\begin{table}
\begin{ruledtabular}
\begin{tabular}{lcdr}
\textrm{Model}&
\textrm{2 state fidelity}& \textrm{3 state fidelity}\\
\colrule
Conventional & 0.841 & 0.711\\
Matched Filter & 0.913 & 0.747 \\
K Nearest Neighbours  &0.902   &0.845\\
Support Vector Machine   &0.917 &0.851\\
Random Forest Classifier   &0.917 &0.874\\ 
Vanilla Neural Network    &0.912 &0.925\\
LSTM &0.909   &0.904\\
CNN  &0.919 &0.928\\
\end{tabular}
\end{ruledtabular}
\caption{Assignment fidelities for different machine learning models evaluated on an identical test data set generated with Sample A. For all methods, there are significant improvements over the conventional method (time integration of readout signal and setting classification thresholds). CNN is the best model for classifying between three levels. Note that the same data was used to extract both the qubit and qutrit fidelities. Since the readout parameters were optimised for the qutrit case the CNN model returns higher value for the 3 state fidelity.}
\label{tab:ml_fid}
\end{table}

We evaluated the assignment fidelity of several machine learning methods: the support vector machine, a random forest classifier, and a k-nearest-neighbors algorithm. After this, we deployed neural network architectures, including a single hidden layer (`vanilla')  feedforward neural network, several versions of convolutional neural networks (CNN), and long short term memory networks (LSTM). The assignment fidelities obtained using each of these algorithms is shown in Table \ref{tab:ml_fid}.

The k-nearest neighbour algorithm bears the most similarity to the matched filter method. The LSTM algorithm is popular in language processing, and was chosen because they are designed to deal with sequences of data, and can therefore process long time correlations between data points.
CNN is most popularly used for pattern recognition and image classification. It is a neural network where the hidden layer is a convolution of the input with a kernel (or filter). We feed the time-domain signal data to the CNN with the I(t) and Q(t) traces as two channel inputs, analogous to the red, blue and green channels of a color image. 

After selecting CNN as the method with the highest assignment fidelity, we apply this model to state classification on-the-fly.  
The network consists of the following layers:
\begin{enumerate}
    \item \textbf{1D convolution:}   A convolution layer with 2 input channels (corresponding to I and Q), 16 output channels, and a kernel size of 128. The large kernel size filters out higher frequency noise effectively. The initial kernel weights (or filter coefficients) are manually set using the He initialisation function \cite{he2015delving}, with experimentation demonstrating that CNN performance is sensitive to this weight initialisation. 
    
    \item \textbf{ReLU activation:} Maps $f(x) = \rm{max}(0,x)$  in order to expedite the learning process \cite{Nair2010}. 
    \item \textbf{1D convolution:} A convolution layer with 16 input channels (corresponding to the output of the previous convolution layer), 32 output channels, and a kernel size of 5. This expands the previous 16 features to 32, by taking various linear combinations of the prior layer outputs.
    \item \textbf{ReLU activation} 
    \item \textbf{Max pooling:} The maximums of every three neighbouring output values of the previous layer are evaluated, to reduce the size of the data representation and therefore computational time in the remainder of the network.
    \item \textbf{Flattening:}  Reshapes data to a one-dimensional array.
    \item \textbf{Dropout:} 50\% of the data points are randomly selected to be set to zero (called `neuron deactivation'). This step was initially introduced to prevent the model from being \textit{overtrained}. Overtraining (or overfitting) is known behaviour when ML methods learns `too much' from the training data, showing an artificially high accuracy of classification while training, but failing to maintain performance when tested on new data. This may occur if there are certain features of the signal present in the training dataset due to noise or other conditions relevant to the data-taking process. 
    \item \textbf{Linear:} Applies a linear transformation to the incoming data: $y= A x +b$, where weights (denoted as a matrix $A$) and biases (denoted as a vector $b$) are optimized. Here, the size of the output $y$ is half of the size of the input $x$.
    \item \textbf{ReLU activation} 
    \item \textbf{Linear:} The size of the output data is designed to be two for the qubit and three for the qutrit classification, corresponding to the possible preparation states.
\end{enumerate}

The Adam optimizer \cite{kingma2017adam} and a mean squared error (MSE) loss function were used for opimization. The output was classified using a softmax function. For each training cycle, we record 2048 traces of each state, and pass these through the model. The loss was calculated and gradient descent was undertaken at a learning rate of $10^{-3}$. This cycle takes $\sim3$ seconds. 

We trained the model on new data at every training cycle, acquired in real-time from the sample. This provides an intrinsic protection from overfitting; since there is a new dataset each time the loss is calculated, the model cannot learn on any spurious signal features localised to a single dataset. The low learning rate assists in helping the model to learn patterns that are common to data across training cycles, thereby increasing model stability. After each update of the model weights, a further 2048 acquisitions were made to test the model. The loss and assignment fidelity on test data were stored for monitoring. 

\begin{figure*}[hbtp]
\centering
\includegraphics[width=\textwidth]{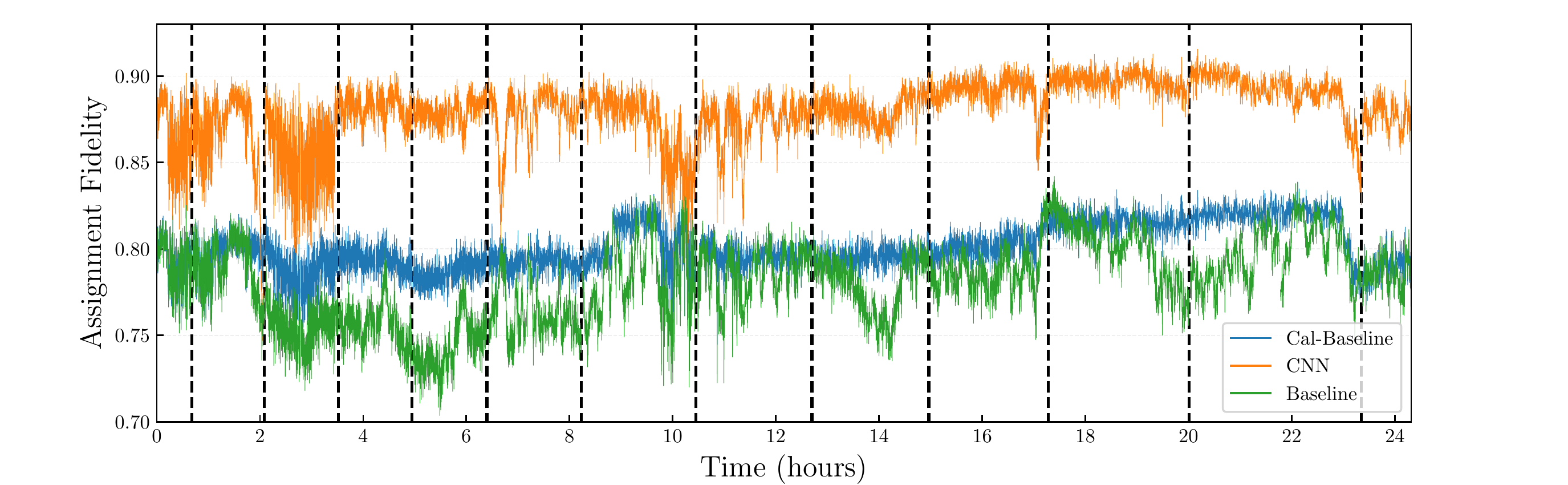}
\caption{Fidelity of different classification methods. Each point represents a fidelity evaluation using 2048 traces for each of the three states gathered on the fly, while each plot represents a different training regime. The black dashed lines indicate when the model was retrained. The CNN model consistently does better than the ``Cal-Baseline'' measurement, even though it is only trained at discrete intervals.}
\label{fig:comparison}
\end{figure*}

To investigate robustness of the CNN classification model against system parameter drifts, we performed continuous measurements over 24 hours and monitored the fidelity values as shown in Fig.~\ref{fig:comparison}. First, we evaluated the fidelities obtained from the integrated responses according to the conventional method (``Baseline"). Second, we obtained the fidelity using the mean responses re-calibrated every 2048 repeated measurements (``Cal-Baseline"). Finally, we plot the fidelity obtained by on-fly-processing with the CNN model. 

\begin{figure}[hbtp]
\centering
\includegraphics[width=\linewidth]{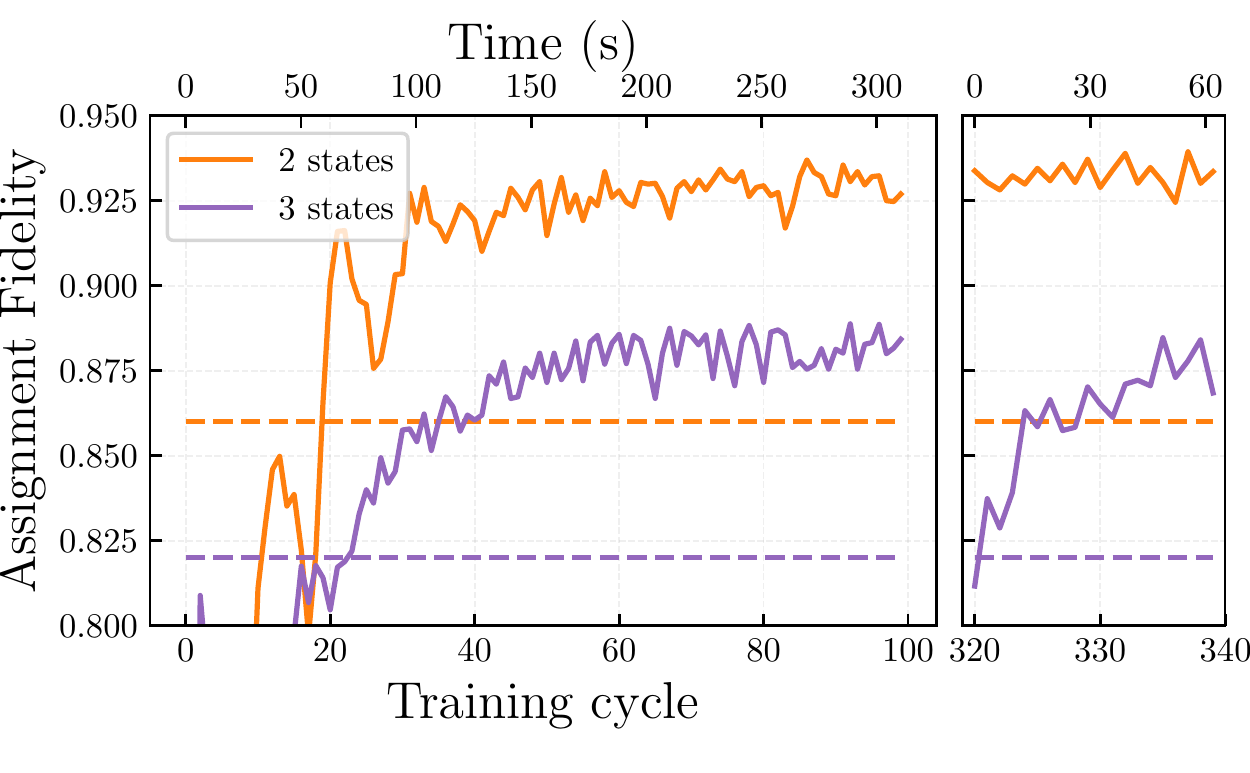}
\caption{Training over 100 training cycles and re-training over 20 cycles. The left plot shows the CNN learning to differentiate between the three (or two) states. The model provides better fidelity than the baseline fidelity within 30 training cycles over approximately one minute. The plot on the right shows the model being re-trained after running the experiment for a few hours. Retraining takes only 20 training cycles.}
\label{fig:training}
\end{figure}

The model is trained for an initial 100 training cycles. The fidelity is then repeatedly tested for $\sim 1$ to $3.5$~hours, before training the model again for another $20$ training cycles. Fig.~\ref{fig:training} shows the assignment fidelity during training and retraining. This retraining process is akin to recalibration in the ``Cal-Baseline'' measurement. This model performs better than both the ``Baseline'' and ``Cal-Baseline'' measurements, despite the frequency of recalibration being significantly lower than its ``Cal-Baseline'' counterpart. Without retraining, the model performs better than the ``Baseline'' throughout the 24 hours, but loses its enhanced fidelity after a few hours.

To enforce robustness to input parameter variations we can train CNN with the dataset containing samples which are broadly distributed across the domain of that input parameter. This effectively removes the parameter from the possible set of features upon which the model can learn. An example of such a parameter variation can be a global phase drift generated by insufficient instrument synchronisation.

To endow the network with global phase robustness, a secondary dataset was created by obtaining 256 traces for each of the three basis states at each of 500 different global phases. Global phase was applied by imposing an arbitrary wait time at the beginning of each experiment repetition. The readout signals obtained within each experiment are downconverted to 25 MHz using a local oscillator which is phase-locked to the generator producing the readout signal itself. This phase-locking ensures that if measurements occur at integer multiples of 40 ns within the pulse sequence, phase coherence is conserved. By generating data with uniformly distributed initial wait times between 0 and 40 ns, a predictable global phase was no longer a feature of the training dataset, forcing the model to learn a phase-robust mapping from readout trajectory to qutrit state. We fed this data into a CNN network in the same manner as the standard dataset.

\begin{figure}[hbtp]
\centering
\includegraphics[width=\linewidth]{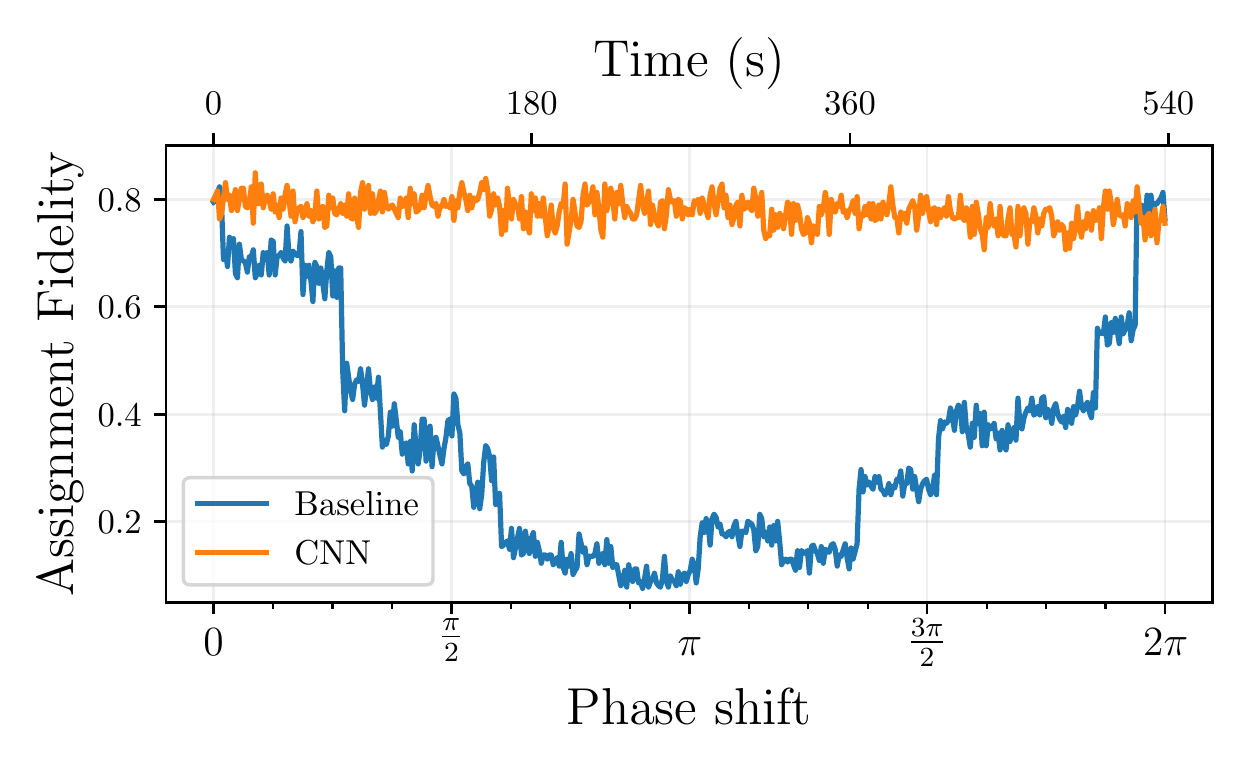}
\caption{Assignment fidelities of the baseline mean calibration method and a phase-robust CNN model classifying data with an induced phase drift. Each point represents a fidelity evaluation using 2048 traces for each of the three states. Neither method undergoes any retraining or recalibration over the course of the experiment.}
\label{fig:phasedrift}
\end{figure}

The design of the neural network remains identical for this experiment, aside from using a kernel size of 10 in the first convolutional layer. Assignment fidelities are evaluated at 500 intervals with phase shifts ranging from 0 to 2$\pi$ over the course of 9 minutes in Fig.~\ref{fig:phasedrift}, using both the baseline mean integration and CNN processing methods (trained/calibrated using data gathered immediately prior). The CNN model classifies with fidelities comparable to the baseline state classification fidelities without phase drift, but any non-trivial phase drift will degrade the conventional approach markedly while the CNN maintains accuracy without any re-calibration required.

In summary, we investigated a set of machine learning methods for classification of a transmon with the dispersive measurement. The method classifies the states when two and three levels of the transmon are considered and can be directly extended to more transmon levels. Some modifications accounting for cross-talk could be necessary for applying this method to multi-qubit systems.
Convolutional neural network methods demonstrated the highest performance consistent with the results from \cite{Lienhard2021}. CNN methods could also be trained against experimental imperfections such as the local oscillator phase drift. Improved assignment fidelity and ability to directly train the model against system imperfections are the key advantages of neural networks. The open-source, GPU-friendly, and easily implementable nature of PyTorch makes these neural networks an attractive tool for state classification.\\

\section*{Supplementary Material}
See supplementary material for the experiment schematic and an example of readout trajectories with and without decay.

\section*{Acknowledgements}
We thank Andreas Wallraff, Markus Oppliger, Anton Poto\v{c}nik, and Mintu Mondal for fabricating JPA used in the measurements. The authors were supported by the Australian Research Council Centre of Excellence for Engineered Quantum Systems (EQUS, CE170100009) and by Lockheed Martin Corporation via research contract S19~004. 

\section*{Data Availability}
The data that support the findings of this study and the code used for generating and training the machine learning algorithm are openly available in UQ eSpace at \nolinkurl{https://doi.org/10.48610/9d3c2eb}. 

\bibliography{MachineLearning}

\begin{thebibliography}{25}%
\makeatletter
\providecommand \@ifxundefined [1]{%
 \@ifx{#1\undefined}
}%
\providecommand \@ifnum [1]{%
 \ifnum #1\expandafter \@firstoftwo
 \else \expandafter \@secondoftwo
 \fi
}%
\providecommand \@ifx [1]{%
 \ifx #1\expandafter \@firstoftwo
 \else \expandafter \@secondoftwo
 \fi
}%
\providecommand \natexlab [1]{#1}%
\providecommand \enquote  [1]{``#1''}%
\providecommand \bibnamefont  [1]{#1}%
\providecommand \bibfnamefont [1]{#1}%
\providecommand \citenamefont [1]{#1}%
\providecommand \href@noop [0]{\@secondoftwo}%
\providecommand \href [0]{\begingroup \@sanitize@url \@href}%
\providecommand \@href[1]{\@@startlink{#1}\@@href}%
\providecommand \@@href[1]{\endgroup#1\@@endlink}%
\providecommand \@sanitize@url [0]{\catcode `\\12\catcode `\$12\catcode
  `\&12\catcode `\#12\catcode `\^12\catcode `\_12\catcode `\%12\relax}%
\providecommand \@@startlink[1]{}%
\providecommand \@@endlink[0]{}%
\providecommand \url  [0]{\begingroup\@sanitize@url \@url }%
\providecommand \@url [1]{\endgroup\@href {#1}{\urlprefix }}%
\providecommand \urlprefix  [0]{URL }%
\providecommand \Eprint [0]{\href }%
\providecommand \doibase [0]{https://doi.org/}%
\providecommand \selectlanguage [0]{\@gobble}%
\providecommand \bibinfo  [0]{\@secondoftwo}%
\providecommand \bibfield  [0]{\@secondoftwo}%
\providecommand \translation [1]{[#1]}%
\providecommand \BibitemOpen [0]{}%
\providecommand \bibitemStop [0]{}%
\providecommand \bibitemNoStop [0]{.\EOS\space}%
\providecommand \EOS [0]{\spacefactor3000\relax}%
\providecommand \BibitemShut  [1]{\csname bibitem#1\endcsname}%
\let\auto@bib@innerbib\@empty
\bibitem [{\citenamefont {Zahedinejad}\ \emph {et~al.}(2016)\citenamefont
  {Zahedinejad}, \citenamefont {Ghosh},\ and\ \citenamefont
  {Sanders}}]{Zahedinejad2016}%
  \BibitemOpen
  \bibfield  {author} {\bibinfo {author} {\bibfnamefont {E.}~\bibnamefont
  {Zahedinejad}}, \bibinfo {author} {\bibfnamefont {J.}~\bibnamefont {Ghosh}},\
  and\ \bibinfo {author} {\bibfnamefont {B.~C.}\ \bibnamefont {Sanders}},\
  }\bibfield  {title} {\bibinfo {title} {Designing high-fidelity single-shot
  three-qubit gates: A machine-learning approach},\ }\href
  {https://doi.org/10.1103/PhysRevApplied.6.054005} {\bibfield  {journal}
  {\bibinfo  {journal} {Phys. Rev. Applied}\ }\textbf {\bibinfo {volume} {6}},\
  \bibinfo {pages} {054005} (\bibinfo {year} {2016})}\BibitemShut {NoStop}%
\bibitem [{\citenamefont {August}\ and\ \citenamefont {Ni}(2017)}]{August2017}%
  \BibitemOpen
  \bibfield  {author} {\bibinfo {author} {\bibfnamefont {M.}~\bibnamefont
  {August}}\ and\ \bibinfo {author} {\bibfnamefont {X.}~\bibnamefont {Ni}},\
  }\bibfield  {title} {\bibinfo {title} {Using recurrent neural networks to
  optimize dynamical decoupling for quantum memory},\ }\href
  {https://doi.org/10.1103/PhysRevA.95.012335} {\bibfield  {journal} {\bibinfo
  {journal} {Phys. Rev. A}\ }\textbf {\bibinfo {volume} {95}},\ \bibinfo
  {pages} {012335} (\bibinfo {year} {2017})}\BibitemShut {NoStop}%
\bibitem [{\citenamefont {Ding}\ \emph {et~al.}(2021)\citenamefont {Ding},
  \citenamefont {Ban}, \citenamefont {Mart\'{\i}n-Guerrero}, \citenamefont
  {Solano}, \citenamefont {Casanova},\ and\ \citenamefont {Chen}}]{Ding2021}%
  \BibitemOpen
  \bibfield  {author} {\bibinfo {author} {\bibfnamefont {Y.}~\bibnamefont
  {Ding}}, \bibinfo {author} {\bibfnamefont {Y.}~\bibnamefont {Ban}}, \bibinfo
  {author} {\bibfnamefont {J.~D.}\ \bibnamefont {Mart\'{\i}n-Guerrero}},
  \bibinfo {author} {\bibfnamefont {E.}~\bibnamefont {Solano}}, \bibinfo
  {author} {\bibfnamefont {J.}~\bibnamefont {Casanova}},\ and\ \bibinfo
  {author} {\bibfnamefont {X.}~\bibnamefont {Chen}},\ }\bibfield  {title}
  {\bibinfo {title} {Breaking adiabatic quantum control with deep learning},\
  }\href {https://doi.org/10.1103/PhysRevA.103.L040401} {\bibfield  {journal}
  {\bibinfo  {journal} {Phys. Rev. A}\ }\textbf {\bibinfo {volume} {103}},\
  \bibinfo {pages} {L040401} (\bibinfo {year} {2021})}\BibitemShut {NoStop}%
\bibitem [{\citenamefont {Baum}\ \emph {et~al.}(2021)\citenamefont {Baum},
  \citenamefont {Amico}, \citenamefont {Howell}, \citenamefont {Hush},
  \citenamefont {Liuzzi}, \citenamefont {Mundada}, \citenamefont {Merkh},
  \citenamefont {Carvalho},\ and\ \citenamefont {Biercuk}}]{Baum2021}%
  \BibitemOpen
  \bibfield  {author} {\bibinfo {author} {\bibfnamefont {Y.}~\bibnamefont
  {Baum}}, \bibinfo {author} {\bibfnamefont {M.}~\bibnamefont {Amico}},
  \bibinfo {author} {\bibfnamefont {S.}~\bibnamefont {Howell}}, \bibinfo
  {author} {\bibfnamefont {M.}~\bibnamefont {Hush}}, \bibinfo {author}
  {\bibfnamefont {M.}~\bibnamefont {Liuzzi}}, \bibinfo {author} {\bibfnamefont
  {P.}~\bibnamefont {Mundada}}, \bibinfo {author} {\bibfnamefont
  {T.}~\bibnamefont {Merkh}}, \bibinfo {author} {\bibfnamefont {A.~R.~R.}\
  \bibnamefont {Carvalho}},\ and\ \bibinfo {author} {\bibfnamefont {M.~J.}\
  \bibnamefont {Biercuk}},\ }\bibfield  {title} {\bibinfo {title} {Experimental
  deep reinforcement learning for error-robust gateset design on a
  superconducting quantum computer},\ }\href {https://arxiv.org/abs/2105.01079}
  {\bibfield  {journal} {\bibinfo  {journal} {arXiv:2105.01079}\ } (\bibinfo
  {year} {2021})}\BibitemShut {NoStop}%
\bibitem [{\citenamefont {Rem}\ \emph {et~al.}(2019)\citenamefont {Rem},
  \citenamefont {K{\"a}ming}, \citenamefont {Tarnowski}, \citenamefont
  {Asteria}, \citenamefont {Fl{\"a}schner}, \citenamefont {Becker},
  \citenamefont {Sengstock},\ and\ \citenamefont {Weitenberg}}]{Rem2019}%
  \BibitemOpen
  \bibfield  {author} {\bibinfo {author} {\bibfnamefont {B.~S.}\ \bibnamefont
  {Rem}}, \bibinfo {author} {\bibfnamefont {N.}~\bibnamefont {K{\"a}ming}},
  \bibinfo {author} {\bibfnamefont {M.}~\bibnamefont {Tarnowski}}, \bibinfo
  {author} {\bibfnamefont {L.}~\bibnamefont {Asteria}}, \bibinfo {author}
  {\bibfnamefont {N.}~\bibnamefont {Fl{\"a}schner}}, \bibinfo {author}
  {\bibfnamefont {C.}~\bibnamefont {Becker}}, \bibinfo {author} {\bibfnamefont
  {K.}~\bibnamefont {Sengstock}},\ and\ \bibinfo {author} {\bibfnamefont
  {C.}~\bibnamefont {Weitenberg}},\ }\bibfield  {title} {\bibinfo {title}
  {Identifying quantum phase transitions using artificial neural networks on
  experimental data},\ }\href {https://doi.org/10.1038/s41567-019-0554-0}
  {\bibfield  {journal} {\bibinfo  {journal} {Nature Physics}\ }\textbf
  {\bibinfo {volume} {15}},\ \bibinfo {pages} {917} (\bibinfo {year}
  {2019})}\BibitemShut {NoStop}%
\bibitem [{\citenamefont {Dong}\ \emph {et~al.}(2019)\citenamefont {Dong},
  \citenamefont {Pollmann},\ and\ \citenamefont {Zhang}}]{Dong2019}%
  \BibitemOpen
  \bibfield  {author} {\bibinfo {author} {\bibfnamefont {X.-Y.}\ \bibnamefont
  {Dong}}, \bibinfo {author} {\bibfnamefont {F.}~\bibnamefont {Pollmann}},\
  and\ \bibinfo {author} {\bibfnamefont {X.-F.}\ \bibnamefont {Zhang}},\
  }\bibfield  {title} {\bibinfo {title} {Machine learning of quantum phase
  transitions},\ }\href {https://doi.org/10.1103/PhysRevB.99.121104} {\bibfield
   {journal} {\bibinfo  {journal} {Phys. Rev. B}\ }\textbf {\bibinfo {volume}
  {99}},\ \bibinfo {pages} {121104} (\bibinfo {year} {2019})}\BibitemShut
  {NoStop}%
\bibitem [{\citenamefont {Palmieri}\ \emph {et~al.}(2020)\citenamefont
  {Palmieri}, \citenamefont {Kovlakov}, \citenamefont {Bianchi}, \citenamefont
  {Yudin}, \citenamefont {Straupe}, \citenamefont {Biamonte},\ and\
  \citenamefont {Kulik}}]{Palmieri2020}%
  \BibitemOpen
  \bibfield  {author} {\bibinfo {author} {\bibfnamefont {A.~M.}\ \bibnamefont
  {Palmieri}}, \bibinfo {author} {\bibfnamefont {E.}~\bibnamefont {Kovlakov}},
  \bibinfo {author} {\bibfnamefont {F.}~\bibnamefont {Bianchi}}, \bibinfo
  {author} {\bibfnamefont {D.}~\bibnamefont {Yudin}}, \bibinfo {author}
  {\bibfnamefont {S.}~\bibnamefont {Straupe}}, \bibinfo {author} {\bibfnamefont
  {J.~D.}\ \bibnamefont {Biamonte}},\ and\ \bibinfo {author} {\bibfnamefont
  {S.}~\bibnamefont {Kulik}},\ }\bibfield  {title} {\bibinfo {title}
  {Experimental neural network enhanced quantum tomography},\ }\href
  {https://doi.org/10.1038/s41534-020-0248-6} {\bibfield  {journal} {\bibinfo
  {journal} {npj Quantum Information}\ }\textbf {\bibinfo {volume} {6}},\
  \bibinfo {pages} {20} (\bibinfo {year} {2020})}\BibitemShut {NoStop}%
\bibitem [{\citenamefont {Zwolak}\ \emph {et~al.}(2020)\citenamefont {Zwolak},
  \citenamefont {McJunkin}, \citenamefont {Kalantre}, \citenamefont {Dodson},
  \citenamefont {MacQuarrie}, \citenamefont {Savage}, \citenamefont {Lagally},
  \citenamefont {Coppersmith}, \citenamefont {Eriksson},\ and\ \citenamefont
  {Taylor}}]{Zwolak2020}%
  \BibitemOpen
  \bibfield  {author} {\bibinfo {author} {\bibfnamefont {J.~P.}\ \bibnamefont
  {Zwolak}}, \bibinfo {author} {\bibfnamefont {T.}~\bibnamefont {McJunkin}},
  \bibinfo {author} {\bibfnamefont {S.~S.}\ \bibnamefont {Kalantre}}, \bibinfo
  {author} {\bibfnamefont {J.}~\bibnamefont {Dodson}}, \bibinfo {author}
  {\bibfnamefont {E.}~\bibnamefont {MacQuarrie}}, \bibinfo {author}
  {\bibfnamefont {D.}~\bibnamefont {Savage}}, \bibinfo {author} {\bibfnamefont
  {M.}~\bibnamefont {Lagally}}, \bibinfo {author} {\bibfnamefont
  {S.}~\bibnamefont {Coppersmith}}, \bibinfo {author} {\bibfnamefont {M.~A.}\
  \bibnamefont {Eriksson}},\ and\ \bibinfo {author} {\bibfnamefont {J.~M.}\
  \bibnamefont {Taylor}},\ }\bibfield  {title} {\bibinfo {title} {Autotuning of
  double-dot devices in situ with machine learning},\ }\href
  {https://doi.org/10.1103/PhysRevApplied.13.034075} {\bibfield  {journal}
  {\bibinfo  {journal} {Phys. Rev. Applied}\ }\textbf {\bibinfo {volume}
  {13}},\ \bibinfo {pages} {034075} (\bibinfo {year} {2020})}\BibitemShut
  {NoStop}%
\bibitem [{\citenamefont {Durrer}\ \emph {et~al.}(2020)\citenamefont {Durrer},
  \citenamefont {Kratochwil}, \citenamefont {Koski}, \citenamefont {Landig},
  \citenamefont {Reichl}, \citenamefont {Wegscheider}, \citenamefont {Ihn},\
  and\ \citenamefont {Greplova}}]{Durrer2020}%
  \BibitemOpen
  \bibfield  {author} {\bibinfo {author} {\bibfnamefont {R.}~\bibnamefont
  {Durrer}}, \bibinfo {author} {\bibfnamefont {B.}~\bibnamefont {Kratochwil}},
  \bibinfo {author} {\bibfnamefont {J.}~\bibnamefont {Koski}}, \bibinfo
  {author} {\bibfnamefont {A.}~\bibnamefont {Landig}}, \bibinfo {author}
  {\bibfnamefont {C.}~\bibnamefont {Reichl}}, \bibinfo {author} {\bibfnamefont
  {W.}~\bibnamefont {Wegscheider}}, \bibinfo {author} {\bibfnamefont
  {T.}~\bibnamefont {Ihn}},\ and\ \bibinfo {author} {\bibfnamefont
  {E.}~\bibnamefont {Greplova}},\ }\bibfield  {title} {\bibinfo {title}
  {Automated tuning of double quantum dots into specific charge states using
  neural networks},\ }\href {https://doi.org/10.1103/PhysRevApplied.13.054019}
  {\bibfield  {journal} {\bibinfo  {journal} {Phys. Rev. Applied}\ }\textbf
  {\bibinfo {volume} {13}},\ \bibinfo {pages} {054019} (\bibinfo {year}
  {2020})}\BibitemShut {NoStop}%
\bibitem [{\citenamefont {Hu}\ \emph {et~al.}(2019)\citenamefont {Hu},
  \citenamefont {Wu}, \citenamefont {Cai}, \citenamefont {Ma}, \citenamefont
  {Mu}, \citenamefont {Xu}, \citenamefont {Wang}, \citenamefont {Song},
  \citenamefont {Deng}, \citenamefont {Zou},\ and\ \citenamefont
  {Sun}}]{Hueaav2761}%
  \BibitemOpen
  \bibfield  {author} {\bibinfo {author} {\bibfnamefont {L.}~\bibnamefont
  {Hu}}, \bibinfo {author} {\bibfnamefont {S.-H.}\ \bibnamefont {Wu}}, \bibinfo
  {author} {\bibfnamefont {W.}~\bibnamefont {Cai}}, \bibinfo {author}
  {\bibfnamefont {Y.}~\bibnamefont {Ma}}, \bibinfo {author} {\bibfnamefont
  {X.}~\bibnamefont {Mu}}, \bibinfo {author} {\bibfnamefont {Y.}~\bibnamefont
  {Xu}}, \bibinfo {author} {\bibfnamefont {H.}~\bibnamefont {Wang}}, \bibinfo
  {author} {\bibfnamefont {Y.}~\bibnamefont {Song}}, \bibinfo {author}
  {\bibfnamefont {D.-L.}\ \bibnamefont {Deng}}, \bibinfo {author}
  {\bibfnamefont {C.-L.}\ \bibnamefont {Zou}},\ and\ \bibinfo {author}
  {\bibfnamefont {L.}~\bibnamefont {Sun}},\ }\bibfield  {title} {\bibinfo
  {title} {Quantum generative adversarial learning in a superconducting quantum
  circuit},\ }\bibfield  {journal} {\bibinfo  {journal} {Science Advances}\
  }\textbf {\bibinfo {volume} {5}},\ \href
  {https://doi.org/10.1126/sciadv.aav2761} {10.1126/sciadv.aav2761} (\bibinfo
  {year} {2019})\BibitemShut {NoStop}%
\bibitem [{\citenamefont {Cimini}\ \emph {et~al.}(2020)\citenamefont {Cimini},
  \citenamefont {Barbieri}, \citenamefont {Treps}, \citenamefont {Walschaers},\
  and\ \citenamefont {Parigi}}]{Cimini2020}%
  \BibitemOpen
  \bibfield  {author} {\bibinfo {author} {\bibfnamefont {V.}~\bibnamefont
  {Cimini}}, \bibinfo {author} {\bibfnamefont {M.}~\bibnamefont {Barbieri}},
  \bibinfo {author} {\bibfnamefont {N.}~\bibnamefont {Treps}}, \bibinfo
  {author} {\bibfnamefont {M.}~\bibnamefont {Walschaers}},\ and\ \bibinfo
  {author} {\bibfnamefont {V.}~\bibnamefont {Parigi}},\ }\bibfield  {title}
  {\bibinfo {title} {Neural networks for detecting multimode wigner
  negativity},\ }\href {https://doi.org/10.1103/PhysRevLett.125.160504}
  {\bibfield  {journal} {\bibinfo  {journal} {Phys. Rev. Lett.}\ }\textbf
  {\bibinfo {volume} {125}},\ \bibinfo {pages} {160504} (\bibinfo {year}
  {2020})}\BibitemShut {NoStop}%
\bibitem [{\citenamefont {Stenberg}\ \emph {et~al.}(2016)\citenamefont
  {Stenberg}, \citenamefont {K\"ohn},\ and\ \citenamefont
  {Wilhelm}}]{Stenberg2016}%
  \BibitemOpen
  \bibfield  {author} {\bibinfo {author} {\bibfnamefont {M.~P.~V.}\
  \bibnamefont {Stenberg}}, \bibinfo {author} {\bibfnamefont {O.}~\bibnamefont
  {K\"ohn}},\ and\ \bibinfo {author} {\bibfnamefont {F.~K.}\ \bibnamefont
  {Wilhelm}},\ }\bibfield  {title} {\bibinfo {title} {Characterization of
  decohering quantum systems: Machine learning approach},\ }\href
  {https://doi.org/10.1103/PhysRevA.93.012122} {\bibfield  {journal} {\bibinfo
  {journal} {Phys. Rev. A}\ }\textbf {\bibinfo {volume} {93}},\ \bibinfo
  {pages} {012122} (\bibinfo {year} {2016})}\BibitemShut {NoStop}%
\bibitem [{\citenamefont {Flurin}\ \emph {et~al.}(2020)\citenamefont {Flurin},
  \citenamefont {Martin}, \citenamefont {Hacohen-Gourgy},\ and\ \citenamefont
  {Siddiqi}}]{Flurin2020}%
  \BibitemOpen
  \bibfield  {author} {\bibinfo {author} {\bibfnamefont {E.}~\bibnamefont
  {Flurin}}, \bibinfo {author} {\bibfnamefont {L.~S.}\ \bibnamefont {Martin}},
  \bibinfo {author} {\bibfnamefont {S.}~\bibnamefont {Hacohen-Gourgy}},\ and\
  \bibinfo {author} {\bibfnamefont {I.}~\bibnamefont {Siddiqi}},\ }\bibfield
  {title} {\bibinfo {title} {Using a recurrent neural network to reconstruct
  quantum dynamics of a superconducting qubit from physical observations},\
  }\href {https://doi.org/10.1103/PhysRevX.10.011006} {\bibfield  {journal}
  {\bibinfo  {journal} {Phys. Rev. X}\ }\textbf {\bibinfo {volume} {10}},\
  \bibinfo {pages} {011006} (\bibinfo {year} {2020})}\BibitemShut {NoStop}%
\bibitem [{\citenamefont {Gentile}\ \emph {et~al.}(2021)\citenamefont
  {Gentile}, \citenamefont {Flynn}, \citenamefont {Knauer}, \citenamefont
  {Wiebe}, \citenamefont {Paesani}, \citenamefont {Granade}, \citenamefont
  {Rarity}, \citenamefont {Santagati},\ and\ \citenamefont
  {Laing}}]{Gentile2021}%
  \BibitemOpen
  \bibfield  {author} {\bibinfo {author} {\bibfnamefont {A.~A.}\ \bibnamefont
  {Gentile}}, \bibinfo {author} {\bibfnamefont {B.}~\bibnamefont {Flynn}},
  \bibinfo {author} {\bibfnamefont {S.}~\bibnamefont {Knauer}}, \bibinfo
  {author} {\bibfnamefont {N.}~\bibnamefont {Wiebe}}, \bibinfo {author}
  {\bibfnamefont {S.}~\bibnamefont {Paesani}}, \bibinfo {author} {\bibfnamefont
  {C.~E.}\ \bibnamefont {Granade}}, \bibinfo {author} {\bibfnamefont {J.~G.}\
  \bibnamefont {Rarity}}, \bibinfo {author} {\bibfnamefont {R.}~\bibnamefont
  {Santagati}},\ and\ \bibinfo {author} {\bibfnamefont {A.}~\bibnamefont
  {Laing}},\ }\bibfield  {title} {\bibinfo {title} {Learning models of quantum
  systems from experiments},\ }\bibfield  {journal} {\bibinfo  {journal}
  {Nature Physics}\ }\href {https://doi.org/10.1038/s41567-021-01201-7}
  {10.1038/s41567-021-01201-7} (\bibinfo {year} {2021})\BibitemShut {NoStop}%
\bibitem [{\citenamefont {Mei}\ \emph {et~al.}(2021)\citenamefont {Mei},
  \citenamefont {Milosavljevic}, \citenamefont {Simpson}, \citenamefont
  {Smetanka}, \citenamefont {Feeney}, \citenamefont {Seguin}, \citenamefont
  {Ha}, \citenamefont {Ha},\ and\ \citenamefont {Reed}}]{Mei2021a}%
  \BibitemOpen
  \bibfield  {author} {\bibinfo {author} {\bibfnamefont {A.~B.}\ \bibnamefont
  {Mei}}, \bibinfo {author} {\bibfnamefont {I.}~\bibnamefont {Milosavljevic}},
  \bibinfo {author} {\bibfnamefont {A.~L.}\ \bibnamefont {Simpson}}, \bibinfo
  {author} {\bibfnamefont {V.~A.}\ \bibnamefont {Smetanka}}, \bibinfo {author}
  {\bibfnamefont {C.~P.}\ \bibnamefont {Feeney}}, \bibinfo {author}
  {\bibfnamefont {S.~M.}\ \bibnamefont {Seguin}}, \bibinfo {author}
  {\bibfnamefont {S.~D.}\ \bibnamefont {Ha}}, \bibinfo {author} {\bibfnamefont
  {W.}~\bibnamefont {Ha}},\ and\ \bibinfo {author} {\bibfnamefont {M.~D.}\
  \bibnamefont {Reed}},\ }\bibfield  {title} {\bibinfo {title} {Optimization of
  quantum-dot qubit fabrication via machine learning},\ }\href
  {https://doi.org/10.1063/5.0040967} {\bibfield  {journal} {\bibinfo
  {journal} {Applied Physics Letters}\ }\textbf {\bibinfo {volume} {118}},\
  \bibinfo {pages} {204001} (\bibinfo {year} {2021})},\ \Eprint
  {https://arxiv.org/abs/https://doi.org/10.1063/5.0040967}
  {https://doi.org/10.1063/5.0040967} \BibitemShut {NoStop}%
\bibitem [{\citenamefont {Magesan}\ \emph {et~al.}(2015)\citenamefont
  {Magesan}, \citenamefont {Gambetta},\ and\ \citenamefont
  {Chow}}]{Magesan2015}%
  \BibitemOpen
  \bibfield  {author} {\bibinfo {author} {\bibfnamefont {E.}~\bibnamefont
  {Magesan}}, \bibinfo {author} {\bibfnamefont {A.~D.}\ \bibnamefont
  {Gambetta}, \bibfnamefont {J.~M. and.~C\'{o}rcoles}},\ and\ \bibinfo {author}
  {\bibfnamefont {J.~M.}\ \bibnamefont {Chow}},\ }\bibfield  {title} {\bibinfo
  {title} {Machine learning for discriminating quantum measurement trajectories
  and improving readout},\ }\href
  {https://doi.org/http://dx.doi.org/10.1103/PhysRevLett.114.200501} {\bibfield
   {journal} {\bibinfo  {journal} {Phys. Rev. Lett.}\ }\textbf {\bibinfo
  {volume} {114}},\ \bibinfo {pages} {200501} (\bibinfo {year}
  {2015})}\BibitemShut {NoStop}%
\bibitem [{\citenamefont {Dickel}\ \emph {et~al.}(2018)\citenamefont {Dickel},
  \citenamefont {Wesdorp}, \citenamefont {Langford}, \citenamefont {Peiter},
  \citenamefont {Sagastizabal}, \citenamefont {Bruno}, \citenamefont {Criger},
  \citenamefont {Motzoi},\ and\ \citenamefont {DiCarlo}}]{Dickel2018}%
  \BibitemOpen
  \bibfield  {author} {\bibinfo {author} {\bibfnamefont {C.}~\bibnamefont
  {Dickel}}, \bibinfo {author} {\bibfnamefont {J.~J.}\ \bibnamefont {Wesdorp}},
  \bibinfo {author} {\bibfnamefont {N.~K.}\ \bibnamefont {Langford}}, \bibinfo
  {author} {\bibfnamefont {S.}~\bibnamefont {Peiter}}, \bibinfo {author}
  {\bibfnamefont {R.}~\bibnamefont {Sagastizabal}}, \bibinfo {author}
  {\bibfnamefont {A.}~\bibnamefont {Bruno}}, \bibinfo {author} {\bibfnamefont
  {B.}~\bibnamefont {Criger}}, \bibinfo {author} {\bibfnamefont
  {F.}~\bibnamefont {Motzoi}},\ and\ \bibinfo {author} {\bibfnamefont
  {L.}~\bibnamefont {DiCarlo}},\ }\bibfield  {title} {\bibinfo {title}
  {Chip-to-chip entanglement of transmon qubits using engineered measurement
  fields},\ }\href {https://doi.org/10.1103/PhysRevB.97.064508} {\bibfield
  {journal} {\bibinfo  {journal} {Phys. Rev. B}\ }\textbf {\bibinfo {volume}
  {97}},\ \bibinfo {pages} {064508} (\bibinfo {year} {2018})}\BibitemShut
  {NoStop}%
\bibitem [{\citenamefont {Kono}\ \emph {et~al.}(2018)\citenamefont {Kono},
  \citenamefont {Koshino}, \citenamefont {Tabuchi}, \citenamefont {Noguchi},\
  and\ \citenamefont {Nakamura}}]{Kono2018}%
  \BibitemOpen
  \bibfield  {author} {\bibinfo {author} {\bibfnamefont {S.}~\bibnamefont
  {Kono}}, \bibinfo {author} {\bibfnamefont {K.}~\bibnamefont {Koshino}},
  \bibinfo {author} {\bibfnamefont {Y.}~\bibnamefont {Tabuchi}}, \bibinfo
  {author} {\bibfnamefont {A.}~\bibnamefont {Noguchi}},\ and\ \bibinfo {author}
  {\bibfnamefont {Y.}~\bibnamefont {Nakamura}},\ }\bibfield  {title} {\bibinfo
  {title} {Quantum non-demolition detection of an itinerant microwave photon},\
  }\href {https://doi.org/10.1038/s41567-018-0066-3} {\bibfield  {journal}
  {\bibinfo  {journal} {Nature Physics}\ } (\bibinfo {year}
  {2018})}\BibitemShut {NoStop}%
\bibitem [{\citenamefont {Martinez}\ \emph {et~al.}(2020)\citenamefont
  {Martinez}, \citenamefont {Rosen},\ and\ \citenamefont
  {DuBois}}]{Martinez2020}%
  \BibitemOpen
  \bibfield  {author} {\bibinfo {author} {\bibfnamefont {L.~A.}\ \bibnamefont
  {Martinez}}, \bibinfo {author} {\bibfnamefont {Y.~J.}\ \bibnamefont
  {Rosen}},\ and\ \bibinfo {author} {\bibfnamefont {J.~L.}\ \bibnamefont
  {DuBois}},\ }\bibfield  {title} {\bibinfo {title} {Improving qubit readout
  with hidden markov models},\ }\href
  {https://doi.org/10.1103/PhysRevA.102.062426} {\bibfield  {journal} {\bibinfo
   {journal} {Phys. Rev. A}\ }\textbf {\bibinfo {volume} {102}},\ \bibinfo
  {pages} {062426} (\bibinfo {year} {2020})}\BibitemShut {NoStop}%
\bibitem [{\citenamefont {Lienhard}\ \emph {et~al.}(2021)\citenamefont
  {Lienhard}, \citenamefont {Veps\"al\"ainen}, \citenamefont {Govia},
  \citenamefont {Hoffer}, \citenamefont {Qiu}, \citenamefont {Rist\`e},
  \citenamefont {Ware}, \citenamefont {Kim}, \citenamefont {Winik},
  \citenamefont {Melville}, \citenamefont {Niedzielski}, \citenamefont {Yoder},
  \citenamefont {Ribeill}, \citenamefont {Ohki}, \citenamefont {Krovi},
  \citenamefont {Orlando}, \citenamefont {Gustavsson},\ and\ \citenamefont
  {Oliver}}]{Lienhard2021}%
  \BibitemOpen
  \bibfield  {author} {\bibinfo {author} {\bibfnamefont {B.}~\bibnamefont
  {Lienhard}}, \bibinfo {author} {\bibfnamefont {A.}~\bibnamefont
  {Veps\"al\"ainen}}, \bibinfo {author} {\bibfnamefont {L.~C.~G.}\ \bibnamefont
  {Govia}}, \bibinfo {author} {\bibfnamefont {C.~R.}\ \bibnamefont {Hoffer}},
  \bibinfo {author} {\bibfnamefont {J.~Y.}\ \bibnamefont {Qiu}}, \bibinfo
  {author} {\bibfnamefont {D.}~\bibnamefont {Rist\`e}}, \bibinfo {author}
  {\bibfnamefont {M.}~\bibnamefont {Ware}}, \bibinfo {author} {\bibfnamefont
  {D.}~\bibnamefont {Kim}}, \bibinfo {author} {\bibfnamefont {R.}~\bibnamefont
  {Winik}}, \bibinfo {author} {\bibfnamefont {A.}~\bibnamefont {Melville}},
  \bibinfo {author} {\bibfnamefont {B.}~\bibnamefont {Niedzielski}}, \bibinfo
  {author} {\bibfnamefont {J.}~\bibnamefont {Yoder}}, \bibinfo {author}
  {\bibfnamefont {G.~J.}\ \bibnamefont {Ribeill}}, \bibinfo {author}
  {\bibfnamefont {T.~A.}\ \bibnamefont {Ohki}}, \bibinfo {author}
  {\bibfnamefont {H.~K.}\ \bibnamefont {Krovi}}, \bibinfo {author}
  {\bibfnamefont {T.~P.}\ \bibnamefont {Orlando}}, \bibinfo {author}
  {\bibfnamefont {S.}~\bibnamefont {Gustavsson}},\ and\ \bibinfo {author}
  {\bibfnamefont {W.~D.}\ \bibnamefont {Oliver}},\ }\bibfield  {title}
  {\bibinfo {title} {Deep neural network discrimination of multiplexed
  superconducting qubit states},\ }\href {https://arxiv.org/abs/2102.12481}
  {\bibfield  {journal} {\bibinfo  {journal} {arXiv:2102.12481}\ } (\bibinfo
  {year} {2021})}\BibitemShut {NoStop}%
\bibitem [{\citenamefont {Paszke}\ \emph {et~al.}(2019)\citenamefont {Paszke},
  \citenamefont {Gross}, \citenamefont {Massa}, \citenamefont {Lerer},
  \citenamefont {Bradbury}, \citenamefont {Chanan}, \citenamefont {Killeen},
  \citenamefont {Lin}, \citenamefont {Gimelshein}, \citenamefont {Antiga},
  \citenamefont {Desmaison}, \citenamefont {Kopf}, \citenamefont {Yang},
  \citenamefont {DeVito}, \citenamefont {Raison}, \citenamefont {Tejani},
  \citenamefont {Chilamkurthy}, \citenamefont {Steiner}, \citenamefont {Fang},
  \citenamefont {Bai},\ and\ \citenamefont {Chintala}}]{NEURIPS2019_9015}%
  \BibitemOpen
  \bibfield  {author} {\bibinfo {author} {\bibfnamefont {A.}~\bibnamefont
  {Paszke}}, \bibinfo {author} {\bibfnamefont {S.}~\bibnamefont {Gross}},
  \bibinfo {author} {\bibfnamefont {F.}~\bibnamefont {Massa}}, \bibinfo
  {author} {\bibfnamefont {A.}~\bibnamefont {Lerer}}, \bibinfo {author}
  {\bibfnamefont {J.}~\bibnamefont {Bradbury}}, \bibinfo {author}
  {\bibfnamefont {G.}~\bibnamefont {Chanan}}, \bibinfo {author} {\bibfnamefont
  {T.}~\bibnamefont {Killeen}}, \bibinfo {author} {\bibfnamefont
  {Z.}~\bibnamefont {Lin}}, \bibinfo {author} {\bibfnamefont {N.}~\bibnamefont
  {Gimelshein}}, \bibinfo {author} {\bibfnamefont {L.}~\bibnamefont {Antiga}},
  \bibinfo {author} {\bibfnamefont {A.}~\bibnamefont {Desmaison}}, \bibinfo
  {author} {\bibfnamefont {A.}~\bibnamefont {Kopf}}, \bibinfo {author}
  {\bibfnamefont {E.}~\bibnamefont {Yang}}, \bibinfo {author} {\bibfnamefont
  {Z.}~\bibnamefont {DeVito}}, \bibinfo {author} {\bibfnamefont
  {M.}~\bibnamefont {Raison}}, \bibinfo {author} {\bibfnamefont
  {A.}~\bibnamefont {Tejani}}, \bibinfo {author} {\bibfnamefont
  {S.}~\bibnamefont {Chilamkurthy}}, \bibinfo {author} {\bibfnamefont
  {B.}~\bibnamefont {Steiner}}, \bibinfo {author} {\bibfnamefont
  {L.}~\bibnamefont {Fang}}, \bibinfo {author} {\bibfnamefont {J.}~\bibnamefont
  {Bai}},\ and\ \bibinfo {author} {\bibfnamefont {S.}~\bibnamefont
  {Chintala}},\ }\bibfield  {title} {\bibinfo {title} {Pytorch: An imperative
  style, high-performance deep learning library},\ }in\ \href
  {http://papers.neurips.cc/paper/9015-pytorch-an-imperative-style-high-performance-deep-learning-library.pdf}
  {\emph {\bibinfo {booktitle} {Advances in Neural Information Processing
  Systems 32}}},\ \bibinfo {editor} {edited by\ \bibinfo {editor}
  {\bibfnamefont {H.}~\bibnamefont {Wallach}}, \bibinfo {editor} {\bibfnamefont
  {H.}~\bibnamefont {Larochelle}}, \bibinfo {editor} {\bibfnamefont
  {A.}~\bibnamefont {Beygelzimer}}, \bibinfo {editor} {\bibfnamefont
  {F.}~\bibnamefont {d\textquotesingle Alch\'{e}-Buc}}, \bibinfo {editor}
  {\bibfnamefont {E.}~\bibnamefont {Fox}},\ and\ \bibinfo {editor}
  {\bibfnamefont {R.}~\bibnamefont {Garnett}}}\ (\bibinfo  {publisher} {Curran
  Associates, Inc.},\ \bibinfo {year} {2019})\ pp.\ \bibinfo {pages}
  {8024--8035}\BibitemShut {NoStop}%
\bibitem [{\citenamefont {Eichler}\ \emph {et~al.}(2014)\citenamefont
  {Eichler}, \citenamefont {Salathe}, \citenamefont {Mlynek}, \citenamefont
  {Schmidt},\ and\ \citenamefont {Wallraff}}]{Eichler2014}%
  \BibitemOpen
  \bibfield  {author} {\bibinfo {author} {\bibfnamefont {C.}~\bibnamefont
  {Eichler}}, \bibinfo {author} {\bibfnamefont {Y.}~\bibnamefont {Salathe}},
  \bibinfo {author} {\bibfnamefont {J.}~\bibnamefont {Mlynek}}, \bibinfo
  {author} {\bibfnamefont {S.}~\bibnamefont {Schmidt}},\ and\ \bibinfo {author}
  {\bibfnamefont {A.}~\bibnamefont {Wallraff}},\ }\bibfield  {title} {\bibinfo
  {title} {Quantum-limited amplification and entanglement in coupled nonlinear
  resonators},\ }\href
  {https://doi.org/http://dx.doi.org/10.1103/PhysRevLett.113.110502} {\bibfield
   {journal} {\bibinfo  {journal} {Phys. Rev. Lett.}\ ,\ \bibinfo {pages}
  {110502}} (\bibinfo {year} {2014})}\BibitemShut {NoStop}%
\bibitem [{\citenamefont {He}\ \emph {et~al.}(2015)\citenamefont {He},
  \citenamefont {Zhang}, \citenamefont {Ren},\ and\ \citenamefont
  {Sun}}]{he2015delving}%
  \BibitemOpen
  \bibfield  {author} {\bibinfo {author} {\bibfnamefont {K.}~\bibnamefont
  {He}}, \bibinfo {author} {\bibfnamefont {X.}~\bibnamefont {Zhang}}, \bibinfo
  {author} {\bibfnamefont {S.}~\bibnamefont {Ren}},\ and\ \bibinfo {author}
  {\bibfnamefont {J.}~\bibnamefont {Sun}},\ }\href@noop {} {\bibinfo {title}
  {Delving deep into rectifiers: Surpassing human-level performance on imagenet
  classification}} (\bibinfo {year} {2015}),\ \Eprint
  {https://arxiv.org/abs/1502.01852} {arXiv:1502.01852 [cs.CV]} \BibitemShut
  {NoStop}%
\bibitem [{\citenamefont {Nair}\ and\ \citenamefont {Hinton}(2010)}]{Nair2010}%
  \BibitemOpen
  \bibfield  {author} {\bibinfo {author} {\bibfnamefont {V.}~\bibnamefont
  {Nair}}\ and\ \bibinfo {author} {\bibfnamefont {G.~E.}\ \bibnamefont
  {Hinton}},\ }\bibfield  {title} {\bibinfo {title} {Rectified linear units
  improve restricted boltzmann machines},\ }in\ \href@noop {} {\emph {\bibinfo
  {booktitle} {Proceedings of the 27th International Conference on
  International Conference on Machine Learning}}},\ \bibinfo {series and
  number} {ICML'10}\ (\bibinfo  {publisher} {Omnipress},\ \bibinfo {address}
  {Madison, WI, USA},\ \bibinfo {year} {2010})\ p.\ \bibinfo {pages}
  {807–814}\BibitemShut {NoStop}%
\bibitem [{\citenamefont {Kingma}\ and\ \citenamefont
  {Ba}(2017)}]{kingma2017adam}%
  \BibitemOpen
  \bibfield  {author} {\bibinfo {author} {\bibfnamefont {D.~P.}\ \bibnamefont
  {Kingma}}\ and\ \bibinfo {author} {\bibfnamefont {J.}~\bibnamefont {Ba}},\
  }\href@noop {} {\bibinfo {title} {Adam: A method for stochastic
  optimization}} (\bibinfo {year} {2017}),\ \Eprint
  {https://arxiv.org/abs/1412.6980} {arXiv:1412.6980 [cs.LG]} \BibitemShut
  {NoStop}%
\end{thebibliography}%

\end{document}


\renewcommand{\thefigure}{S\arabic{figure}}

\title{Supplementary material - Neural networks for on-the-fly single-shot state classification}

\author{Rohit Navarathna$^{1,2}$}
\email{r.navarathna@uq.edu.au}
\author{Tyler Jones$^{1,2,3}$}
\author{Tina Moghaddam$^{1,2}$}
\author{Anatoly Kulikov$^{1,2,4}$}
\author{Rohit Beriwal$^{1,2}$}
\author{Markus Jerger$^{1,2,5}$}
\author{Prasanna Pakkiam$^{1,2}$}
\author{Arkady Fedorov$^{1,2}$}
\affiliation{
 $^1$ ARC Centre of Excellence for Engineered Quantum Systems, St Lucia, Queensland 4072, Australia\\
 $^2$ School of Mathematics and Physics, University of Queensland, St Lucia, Queensland 4072, Australia\\
 $^3$ Max Kelsen, Spring Hill, Queensland 4000, Australia\\
 $^4$ Department of Physics, ETH Z\"{u}rich, CH-8093 Z\"{u}rich, Switzerland\\
 $^5$ JARA-FIT Institute for Quantum Information, Forschungszentrum J\"{u}lich, 52425 J\"{u}lich, Germany
}
\date{\today}

\maketitle

In this supplementary material, we provide a schematic of the experimental setup used (Fig. 1) and present contrasting readout trajectories demonstrating decay during measurement (Fig. 2). \\

\begin{figure*}[hbtp]
\centering
\includegraphics[width=.83\linewidth]{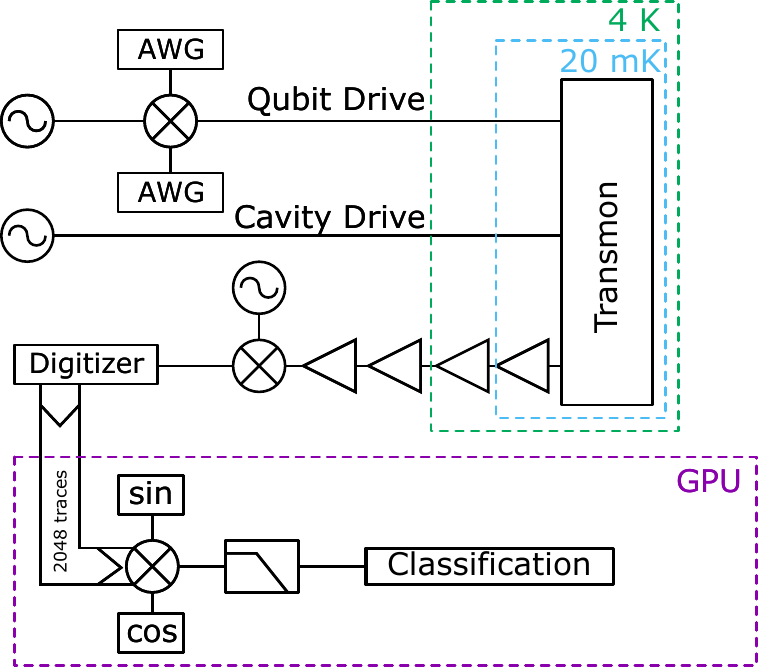}
\caption{Experimental setup. The cavity drive is generated by a pulsed microwave signal generator. The transmon drive uses an upconversion scheme with two AWG channels and a microwave signal generator leading to the charge line of the transmon. The transmission through the cavity is amplified using a JPA, HEMT and room temperature amplifiers. The signal is then downconverted to 25 MHz using a mixer and digitized. The digitized traces are sent to the GPU in batches, then downconverted to DC and filtered using an FIR low pass filter. This is the data which is processed and classified.}
\label{fig:schematic}
\end{figure*}

\begin{figure*}[hbtp]
\centering
\includegraphics[width=\linewidth]{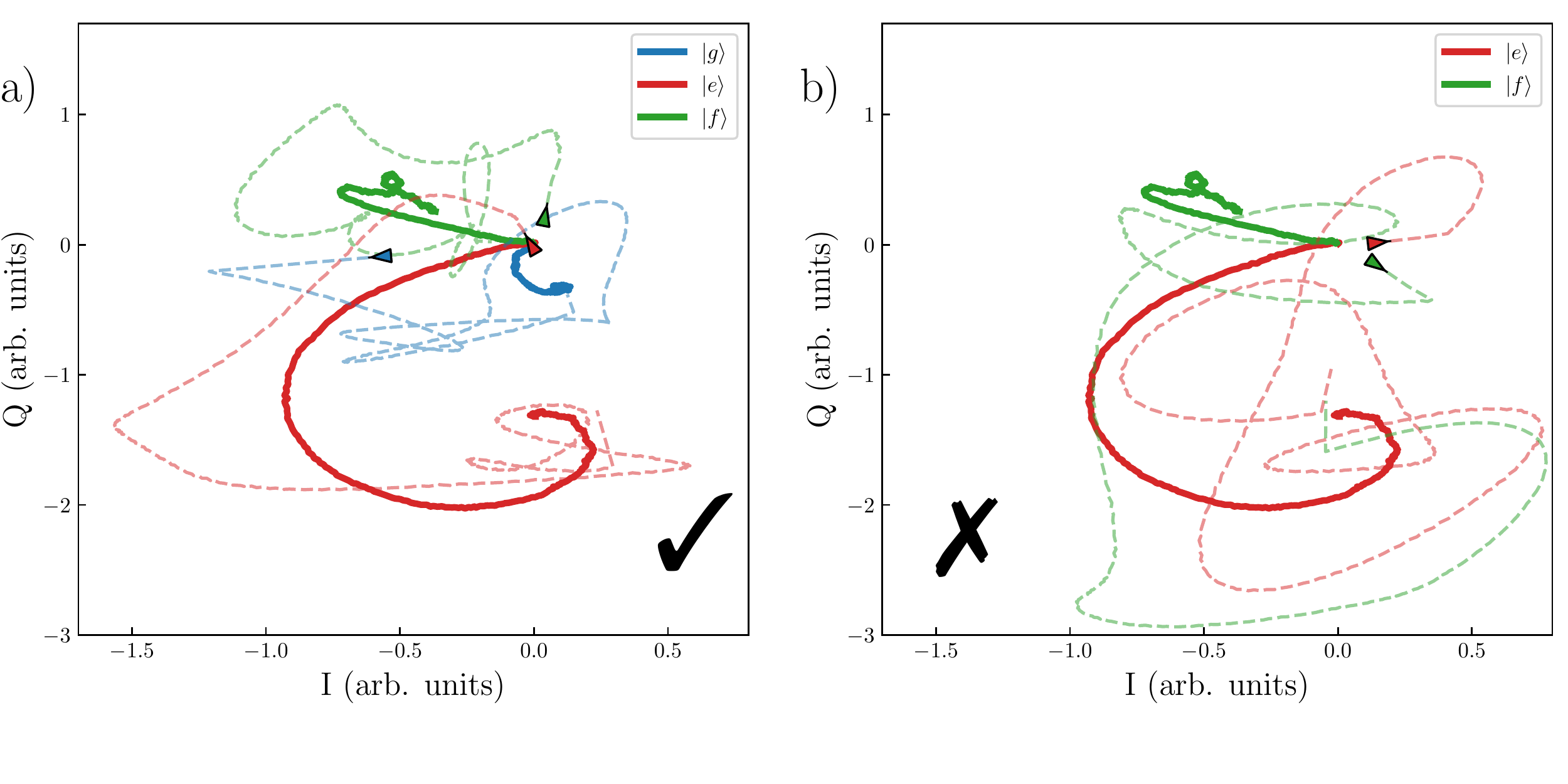}
\caption{Trajectories for the three states. The solid lines are the mean of 2048 traces acquired for each prepared state. The dashed lines are examples of single shots, which are the inputs to the machine learning model. The arrows indicate the direction of the single shot trajectories. The mean trajectories start from $(\rm{I},\rm{Q}) = (0,0)$. a) Both the conventional and CNN correctly classify the shots. b) The CNN correctly classifies all shots, but the conventional method wrongly classifies the $\left|f\right>$ shot as $\left|e\right>$. This is likely because the transmon decayed from $\left|f\right>$ to $\left|e\right>$ early in its trajectory.}
\label{fig:trajectories}
\end{figure*}